\documentclass[journal,transmag]{IEEEtran}
\usepackage{amssymb,mathtools,graphicx,color}
\usepackage{array}
\usepackage{multirow}
\usepackage{booktabs}
\usepackage{hyperref}
\usepackage{silence}
\usepackage{lipsum}
\usepackage{color,soul}
\usepackage{bm}
\usepackage[export]{adjustbox}
\usepackage[linesnumbered,algoruled,boxed,lined]{algorithm2e}
\ifCLASSOPTIONcompsoc
  \usepackage[nocompress]{cite}
\else
  \usepackage{cite}
\fi
\ifCLASSINFOpdf
\else
\fi
\begin{document}
%
% paper title
% Titles are generally capitalized except for words such as a, an, and, as,
% at, but, by, for, in, nor, of, on, or, the, to and up, which are usually
% not capitalized unless they are the first or last word of the title.
% Linebreaks \\ can be used within to get better formatting as desired.
% Do not put math or special symbols in the title.
\title{Spatio-Temporal Segmentation in 3D Echocardiographic \\Sequences using Fractional Brownian Motion}
%
%
% author names and IEEE memberships
% note positions of commas and nonbreaking spaces ( ~ ) LaTeX will not break
% a structure at a ~ so this keeps an author's name from being broken across
% two lines.
% use \thanks{} to gain access to the first footnote area
% a separate \thanks must be used for each paragraph as LaTeX2e's \thanks
% was not built to handle multiple paragraphs
%
%
%\IEEEcompsocitemizethanks is a special \thanks that produces the bulleted
% lists the Computer Society journals use for "first footnote" author
% affiliations. Use \IEEEcompsocthanksitem which works much like \item
% for each affiliation group. When not in compsoc mode,
% \IEEEcompsocitemizethanks becomes like \thanks and
% \IEEEcompsocthanksitem becomes a line break with idention. This
% facilitates dual compilation, although admittedly the differences in the
% desired content of \author between the different types of papers makes a
% one-size-fits-all approach a daunting prospect. For instance, compsoc 
% journal papers have the author affiliations above the "Manuscript
% received ..."  text while in non-compsoc journals this is reversed. Sigh.

\author{Omar S. Al-Kadi,~\IEEEmembership{Senior Member,~IEEE}% <-this % stops a space
\IEEEcompsocitemizethanks{\IEEEcompsocthanksitem O. S. Al-Kadi is with King Abdullah II School for Information Technology, University of Jordan, Amman 11942, Jordan; and also with the Department of Radiology \& Biomedical Imaging, Yale University, New Haven, CT, 06520.\protect\\
% note need leading \protect in front of \\ to get a newline within \thanks as
% \\ is fragile and will error, could use \hfil\break instead.
E-mail: o.alkadi@ju.edu.jo; omar.al-kadi@yale.edu}% <-this % stops a space
}% \thanks{Manuscript received April 19, 2005; revised August 26, 2015.}}

% note the % following the last \IEEEmembership and also \thanks - 
% these prevent an unwanted space from occurring between the last author name
% and the end of the author line. i.e., if you had this:
% 
% \author{....lastname \thanks{...} \thanks{...} }
%                     ^------------^------------^----Do not want these spaces!
%
% a space would be appended to the last name and could cause every name on that
% line to be shifted left slightly. This is one of those "LaTeX things". For
% instance, "\textbf{A} \textbf{B}" will typeset as "A B" not "AB". To get
% "AB" then you have to do: "\textbf{A}\textbf{B}"
% \thanks is no different in this regard, so shield the last } of each \thanks
% that ends a line with a % and do not let a space in before the next \thanks.
% Spaces after \IEEEmembership other than the last one are OK (and needed) as
% you are supposed to have spaces between the names. For what it is worth,
% this is a minor point as most people would not even notice if the said evil
% space somehow managed to creep in.

% The paper headers
\markboth{Accepted manuscript in IEEE Transactions on Biomedical Engineering, Dec 2019. \href{http://dx.doi.org/10.1109/TBME.2019.2958701}{\MakeLowercase{http://dx.doi.org/10.1109/TBME.2019.2958701}}}.%
\IEEEtitleabstractindextext{%
\begin{abstract}
An important aspect for an improved cardiac functional analysis is the accurate segmentation of the left ventricle (LV). A novel approach for fully-automated segmentation of the LV endocardium and epicardium contours is presented. This is mainly based on the natural physical characteristics of the LV shape structure. Both sides of the LV boundaries exhibit natural elliptical curvatures by having details on various scales, i.e. exhibiting \textit{fractal-like} characteristics. The fractional Brownian motion (fBm), which is a non-stationary stochastic process, integrates well with the stochastic nature of ultrasound echoes. It has the advantage of representing a wide range of non-stationary signals and can quantify statistical local self-similarity throughout the time-sequence ultrasound images. The locally characterized boundaries of the fBm segmented LV were further iteratively refined using global information by means of second-order moments. The method is benchmarked using synthetic 3D+time echocardiographic sequences for normal and different ischemic cardiomyopathy, and results compared with state-of-the-art LV segmentation. Furthermore, the framework was validated against real data from canine cases with expert-defined segmentations and demonstrated improved accuracy. The fBm-based segmentation algorithm is fully automatic and has the potential to be used clinically together with 3D echocardiography for improved cardiovascular disease diagnosis.
\end{abstract}

% Note that keywords are not normally used for peerreview papers.
\begin{IEEEkeywords}
fractional Brownian motion, 3D echocardiography, left ventricle, segmentation, heterogeneity, endocardium
\end{IEEEkeywords}}

% make the title area
\maketitle

% To allow for easy dual compilation without having to reenter the
% abstract/keywords data, the \IEEEtitleabstractindextext text will
% not be used in maketitle, but will appear (i.e., to be "transported")
% here as \IEEEdisplaynontitleabstractindextext when compsoc mode
% is not selected <OR> if conference mode is selected - because compsoc
% conference papers position the abstract like regular (non-compsoc)
% papers do!
\IEEEdisplaynontitleabstractindextext
% \IEEEdisplaynontitleabstractindextext has no effect when using
% compsoc under a non-conference mode.

% For peer review papers, you can put extra information on the cover
% page as needed:
% \ifCLASSOPTIONpeerreview
% \begin{center} \bfseries EDICS Category: 3-BBND \end{center}
% \fi
%
% For peerreview papers, this IEEEtran command inserts a page break and
% creates the second title. It will be ignored for other modes.
\IEEEpeerreviewmaketitle

\ifCLASSOPTIONcompsoc
\IEEEraisesectionheading{\section{Introduction}\label{sec:introduction}}
\else
\section{Introduction}
\label{sec:introduction}
\fi
% Computer Society journal (but not conference!) papers do something unusual
% with the very first section heading (almost always called "Introduction").
% They place it ABOVE the main text! IEEEtran.cls does not automatically do
% this for you, but you can achieve this effect with the provided
% \IEEEraisesectionheading{} command. Note the need to keep any \label that
% is to refer to the section immediately after \section in the above as
% \IEEEraisesectionheading puts \section within a raised box.

% The very first letter is a 2 line initial drop letter followed
% by the rest of the first word in caps (small caps for compsoc).
% 
% form to use if the first word consists of a single letter:
% \IEEEPARstart{A}{demo} file is ....
% 
% form to use if you need the single drop letter followed by
% normal text (unknown if ever used by the IEEE):
% \IEEEPARstart{A}{}demo file is ....
% 
% Some journals put the first two words in caps:
% \IEEEPARstart{T}{his demo} file is ....
% 
% Here we have the typical use of a "T" for an initial drop letter
% and "HIS" in caps to complete the first word.
\IEEEPARstart{T}{he} automatic segmentation of the left ventricle (LV) of the heart is still considered an open challenge in the field of medical image segmentation. The interest in understanding this largest and main pumping chamber of the heart refers to the key role it plays in blood circulation. Clinicians mainly assess the extent of heart muscle damage by measuring the LV ejection fraction \cite{lai16}. The accurate and reliable segmentation of the LV shape, and being able to early characterize ischemic myocardial damage, is an important prerequisite for further quantitative analysis of cardiac function.

In clinical practice, the segmentation task involves the delineation of LV endocardium and epicardium contours. This process, however, is tedious and time consuming and prone to intra- and inter-observer variability \cite{nbl06}. An automatic and robust approach for segmenting cardiac ultrasound time-sequences would highly facilitate the routine clinical work \cite{frg01}. Several key challenges in the automated segmentation of the LV in cardiac ultrasound datasets exist. Namely, \textit{speckle intensity heterogeneity}: as in LV cavity (blood pool) due to blood flow or the dynamic motion of the heart; \textit{obscureness}: close proximity of the papillary muscles tend to show speckle intensities similar to that of the myocardium, and thus affecting endocardium segmentation;  \textit{spatial complexity of anatomy}: the separating border between right and left ventricle, and the low contrast between the myocardium and lung air makes segmentation of the epicardium especially difficult. Other factors as partial volume effects due to limited resolution and inter-variability in shape and speckle intensity of the heart chambers across patients due to pathology may pose additional challenges for LV segmentation.
To address these challenges, it is advantageous to have an efficient segmentation algorithm that is objective and reproducible to accelerate and facilitate the process of diagnosis. Regarding ultrasound B-mode segmentation efforts, the endocardium and epicardium boundaries are delineated using a variety of strategies. In particular, statistical models which encode high-level knowledge, as the parametric or geometric deformable models, can handle topological changes robustly \cite{vrg16,bkf16, stg15,cro13}. However, the boundary finding process requires a model to be initialized sufficiently close to the object to converge and is sometimes prone to local minima. Also, many machine learning approaches were proposed for voxel-based classification \cite{dmg14, brd16}, yet the feature engineering process is not straightforward and computationally expensive. Taking advantage of labelled data for detecting data-driven features has drawn increased attention in prior probabilistic maps \cite{hns14,pdr17} and deep learning techniques \cite{cro12,oky18}. The unsupervised learning by deep networks can reduce the need for feature engineering -- one of the most time-consuming parts of machine learning practice, and have recently shown very promising results for improving image classification and segmentation. However, limited training data is a common obstacle in the latter techniques, and regularization of the training data with a large amount of human-annotated data or anatomical models from large datasets is not always available. A review on ultrasound image segmentation methods with techniques mainly focusing on B-mode images can be found in \cite{nbl06}. An intuitive approach would be to incorporate the spatio-temporal domain for improving structure and inter-dependencies of the output.
Dealing with the LV segmentation problem from a spatio-temporal perspective can give further information on the shape boundaries. Due to the nature of the speckle pattern, it is hard to draw conclusions about the boarder of the LV from still frames. Thus, cardiologists usually examine videos of the deformation of the LV wall during the echocardiographic examination. It is logical to assume the speckle pattern structure is better localized when the spatio-temporal coherence is considered. In this regard, Huang et al. exploits the spatio-temporal coherence of individual data for cardiac contour estimation \cite{hng14}. Others embarked on introducing temporal consistency in the tracked speckle pattern using optical flow \cite{sdr17} or gradient vector flow approach \cite{zgk05}. Nevertheless, the spatio-temporal structures of the speckle patterns are usually stochastic in nature. Neglecting the inherent heterogeneity might not characterize the structure in space and time efficiently. The fractional Brownian motion (fBm), which is a non-stationary stochastic process, integrates well with the stochastic nature of ultrasound echoes \cite{alk15}. It has the advantage of representing a wide range of non-stationary signals and can quantify statistical self-similarity in time-sequence ultrasound images. 

To address the aforementioned challenges, we present a novel physically motivated stochastic model for improved epicardium and endocardium boundary segmentation. To our knowledge, this is the first time a spatio-temporal fBm process is used for LV segmentation in 3D ultrasound sequences. Motion roughness, reflected in the speckle pattern, is characterized by fBm for local boundary delineation -- which is theoretically invariant to intensity transformations \cite{alk17}. Using second-order moments for LV shape global information complements the local characterization of the fBm process.

\begin{figure}
 \begin{center}
 \includegraphics[scale=0.30]{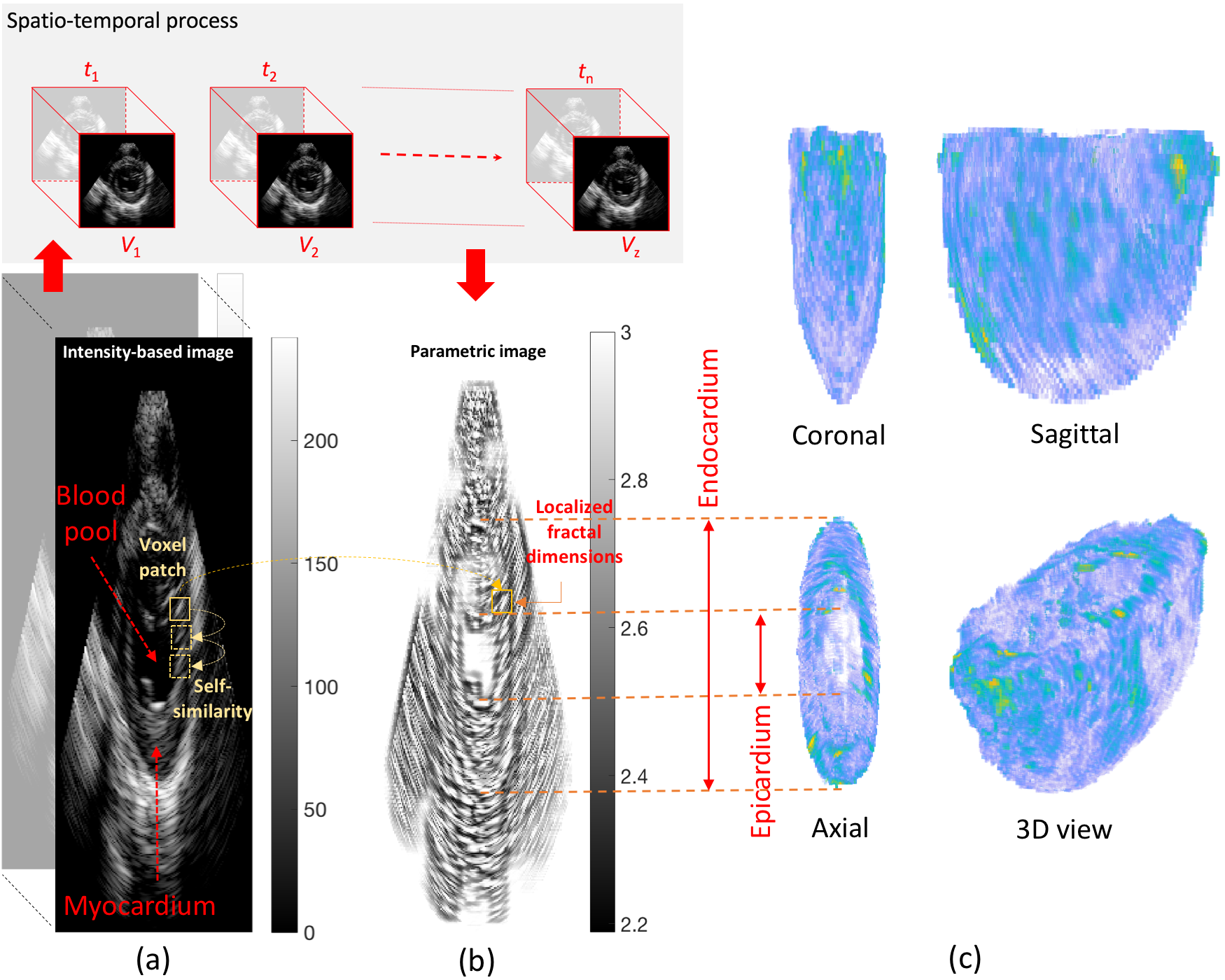}
 \end{center}
\caption{B-mode cardiac ultrasound image of a canine subject showing (a) the left ventricle, (b) corresponding fractional Brownian motion surface image, and (c) 3D reconstruction and projections in different planes. Individual voxels with green-like color represent radial strain during end-systole.}
\label{fig:fBm_surf}
\end{figure}

The direct delineation of the endocardium boundaries is challenging, even for the trained eye, where poor edge information result in some regions becoming obscured, as in the case of the upper left region of Fig.~\ref{fig:fBm_surf}(a). The 3D shape modeling using surface parameterization in Fig.~\ref{fig:fBm_surf}(b) naturally reflects the physics of the LV wall motion and clearly highlights the blood pool and its associated myocardium boundary. The similarity between speckle patterns is evaluated by analyzing their statistical behavior, in particular, by measuring voxel-by-voxel the distance between their statistical distributions. Context around each voxel of interest is found and optimized in a pair-wise pattern detection approach. The fBm herein serves as a similarity criterion to identify such patterns. The self-similarity between the endocardium voxel patches -- quantified as changes in specular reflections -- appear more prominent in the generated fBm parametric image. Similarly applies to the region inside the left ventricular cavity where the speckle patterns are considered more homogeneous. By incorporating the spatio-temporal coherence, an effective segmentation can be achieved. The derived fractal dimensions would reflect the spatio-temporal coherence of the LV boundaries that exhibit natural elliptical curvatures. In this context, the fBm process would adapt to both myocardium and ventricular cavity structures that have sufficient spatio-temporal coherence of varying scale speckle patterns.

In this paper, we introduce a fully-automatic method for robustly segmenting the endocardium and epicardium surface from 3D cardiac ultrasound sequences. In Section~\ref{method}, a 3D stochastic fractal model approach using surface parameterization is adopted for capturing the complex anatomical structure of the LV (see subsection~\ref{method_fBm}). Both local and global information of the LV shape boundaries are characterized with improved precision (see subsection~\ref{method_mont}). Experimental results are reported in Section~\ref{results}. Interpretation and analysis of the results -- with suggestions to future directions, and summary of major findings are discussed in Sections~\ref{discuss} and~\ref{conclude}, respectively.

\section{Methodology}
\label{method}
This sections describes a method for improving LV segmentation, namely, endocardium and epicardium boundaries in 3D cardiac ultrasound time series data. 
%\subsection{Framework overview} 
%\label{overview}
%An integrated fourfold framework is proposed for improving the segmentation of the LV endocardium and epicardium contours. In the first stage, the stochastic model is initialized in a fully-automatic approach based on centroid localization of the basal LV slice. Then the heterogeneity in the structure of the LV time series motion is estimated using fBm. The weights of the spatio-temporal fBm estimation result in a 3D map of localized fractal dimensions, which are used to derive several heterogeneity features from local patches. Then a structured Bayesian classifier is used to predict the probability of each voxel belonging to the epicardium/endocardium-blood interface from given image patches. Following detection of the myocardium surface, an explicit elliptical LV surface model based on image moments was fitted to a subset of the boundary candidates in each volume, in a standard model-to-data approach, where all detected boundary candidates were used to fit the surface.
\subsection{Automatic Initialization}
\label{auto_init}
Formally, let $\bm{\Gamma} \times \bm{\tau} \subset \mathbb{R}^2 \times \mathbb{R}$ be the acquisition space-time of the reference time-series of images $\bm{I}$, $\bm{I} : \bm{\Gamma} \times \bm{\tau} \to \mathbb{R}$, $(x, t) \mapsto \bm{I}(x, t)$. The fBm segmentation maps $\bm{\mathcal{F}}$, representing the fractal dimension computed for each voxel in each 2D slice (derived in subsection (\ref{method_fBm})), is a combination of a spatial transformation $\bm{x^\prime} = \bm{\mathcal{F}}_{space}(x, t)$ and a temporal transformation $\bm{t^\prime} = \bm{\mathcal{F}}_{time}(x, t)$. We make the reasonable assumption that the temporal transformation $\bm{\mathcal{F}}_{time}$ is only time dependent : $\bm{t^\prime} = \bm{\mathcal{F}}_{time}(t)$.
The initialization method only requires to store a database with sample patches describing the possible variations found in LV ventricular cavity and myocardium. The left ventricular cavity is divided into a stack of ellipsoidal-\textit{like} discs (conducted on average $\sim35$ images per sequence). A two-step postprocessing stage is applied to local information: the first step is aimed at filling voxel gaps in segmented ventricular cavity, while the second step is aimed at removing falsely detected isolated myocardium voxels, cf. subsection~\ref{psz_locfBm}. Techniques such as thresholding, mathematical morphology and correlation are combined for this purpose. Then the method entails the identification of base slice centroid acting as a reference for the fBm segmented LV on 3D echocardiography over time. The centroid is tracked throughout the entire fBm segmented left ventricular cavity, where the local shape and orientation varies as the elliptical model parameters are computed towards the apex. This is iterated twice for the endocardium and epicardium boundaries. The volume of the left ventricular cavity represents the sum of the volumes of each of these disks. The volume of each disc is calculated as the cross-sectional area of the disc multiplied by its height.

\begin{figure}
 \begin{center}
 \includegraphics[scale=0.29]{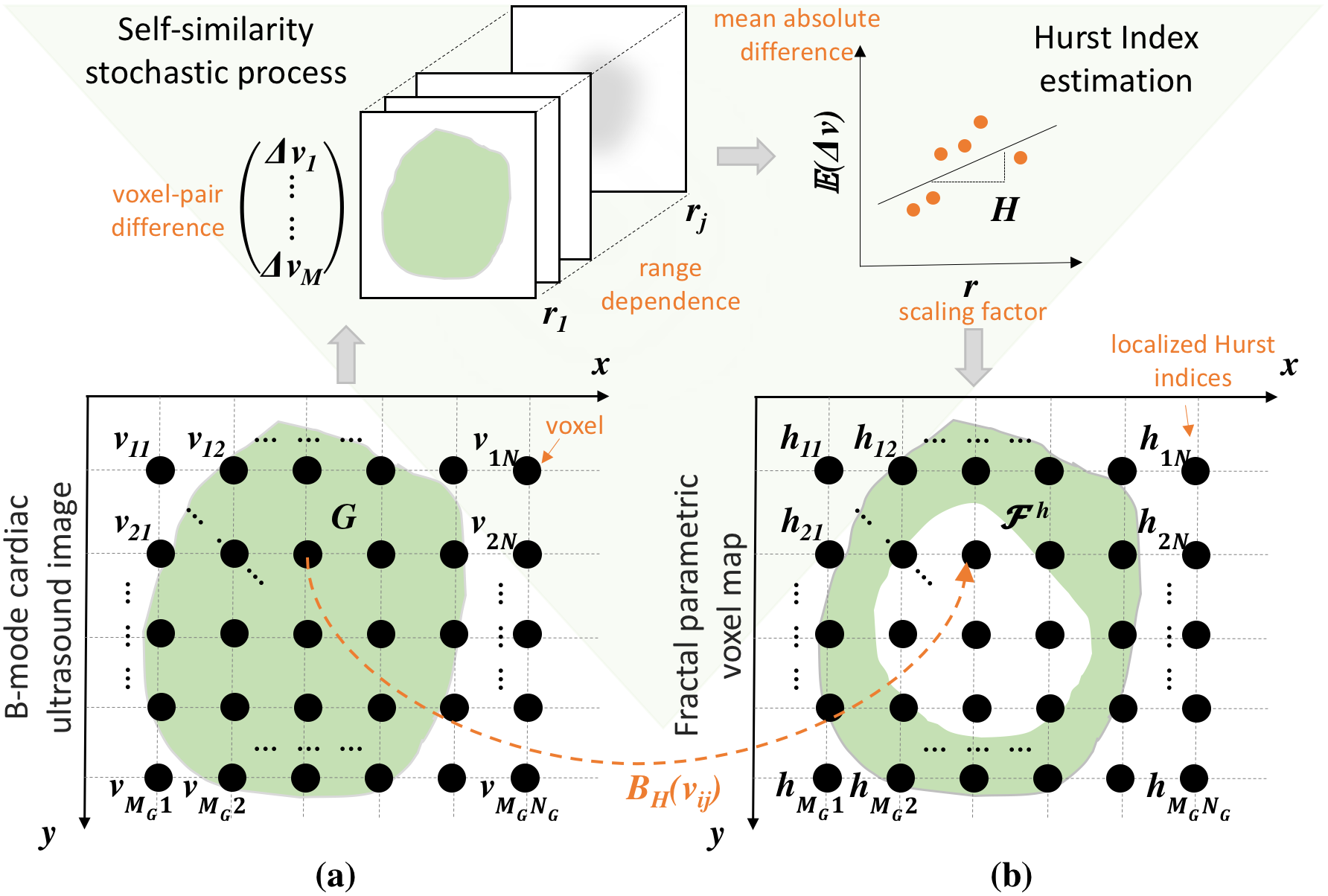}
 \end{center}
\caption{Multiscale parametric mapping based on fractional Brownian motion ($\tilde{B}_{H}$) for voxel-based segmentation. (a) Object $G$ having a spatial support $v_{1,1} \times v_{m_G,n_G}$ in source image $f_{x,y}$, and (b) constructed fractal parametric voxel map $\mathcal{F}^h$ with scaling factor $r$ at different scales ($j$), s.t. each parametric value represents a localized Hurst index ($h$).}
\label{fig:paramap_hurst}
\end{figure}

\subsection{Fractional Brownian Motion}
\label{method_fBm}
Brownian motion $B(t)$ is a \textit{Markovian} process, whose conditional transition density function is time-homogeneous. That is, the probability $\mathbb{P}$ of being in state $B(t)$ at time $t$, given all states up to time $t - 1$, depends only on the previous state, $B(t - 1)$, at time $t - 1$. Therefore, the prediction of what states will occur in the future depends only on the current state. Also, the standard $B(t)$ is a Gaussian process, since it has a normal distribution with specific first two moments. Based on the Lagrangian representation, the generalization of a standard $B(t)$ is a Fractional Brownian motion $B_{H}(t)$, which is a continuous Gaussian self-similar process in $\mathbb{R}$ with stationary increments \cite{MandelbrotNess1968}. $B_{H}(t)$ can be modeled via stochastic integral equation, given by

\begin{gather}
B_{H}(t) - B_{H}(0) = \frac{1}{\Gamma (H + \tfrac{1}{2})}\bigg\{\int\limits_{-\infty}^{0}(t - s)^{H - 1/2} - \nonumber\\(-s)^{H - 1/2} dB(s)+ \int\limits_{0}^{t}(t - s)^{H - 1/2} dB(s) \bigg\}\label{eqn:cont_fBm},
\end{gather}

\noindent where $\Gamma (x)$ and $B(t)$ are the gamma function and standard Brownian motion, respectively, and $H \in (0,1)$ is called the $Hurst$ parameter or index which describes the scaling behavior of the process \cite{mly12} and the roughness of the resultant motion or trajectory \cite{mdl04}. For our case, $H$ characterizes the deformation of the left ventricle wall motion, with lower values leading to a heterogeneous motion and vice versa. From (\ref{eqn:cont_fBm}), the standard Brownian motion (also often called Wiener process) is recovered when $H = \tfrac{1}{2}$. But in contrast to Brownian motion, fBm has dependent increments when $H \neq \tfrac{1}{2}$. By allowing $H$ to differ from $\tfrac{1}{2}$, a fBm process is achieved, where for $H >\frac{1}{2}$ increments are positive correlated, and for $H < \frac{1}{2}$ increments are negatively correlated. The subtracted term $(-s)^{H - 1/2}$ allows the kernel to vanish quickly when $s \to -\infty$, ensuring the convergence of the integral in the range $ 0 < H < 1$.

There are several ways to estimate the fractal dimension of a stochastic process modeled by a discrete-time representation of fractional Brownian motion $\tilde{B}_{H}$ \cite{alk17}. All of them are based on the formula
\begin{equation}
    \label{eqn:var_fBm}
    \mathbb{E}\bigg[|\tilde{B}_{H}(n+l)-\tilde{B}_{H}(n)| \bigg] = c|l|^{H},
\end{equation}
where $c$ is proportional to the standard deviation $\sigma$ for $l$ samples apart. The fBm and power-law variogram fits were used to estimate $H$ as a measure of self-similarity. Particularly, the linearly-related $H$ and corresponding fractal dimension can be calculated from the slope of the average absolute difference plotted as a function of the increments $n$ with sampling interval (or step-size) $l$ on a $\log$-$\log$ plot. Since $H$ is a real number with values varying between a random walk process $H=0$ and a smooth process ($H=1$), the sampling interval can be divided by any arbitrary positive value $l$ and the result rescaled in the ratio $l^H$, cf. property 3 in Appendix \ref{appendix}. Then the new semivariogram will be identical to the initial one. In this sense $\tilde{B}_{H}$ is a self-similar or fractal process. Fig.~\ref{fig:paramap_hurst} illustrates the process of constructing a fractal parametric voxel map $\mathcal{F}^h$ for an object $G$, representing here the blood pool and associated myocardium boundary, from a source image $f_{x,y}$. The resolution for each image slice is investigated for self-similarity patterns by probing for higher resolutions, i.e. searching for voxel pair structures $\Delta v$ that exhibit self-similarity at different scales in the range $r = 1 \dots j$, where $j$ represents the maximum probed scale. Then the variogram estimates the mean absolute difference of each voxel pair $E(\Delta v)$ to the scaling factor $r$, such that the resulting localized Hurst indices $\{h_{11},h_{12},h_{13}, \dots$, $h_{mn}\} \in \mathcal{F}^h$.

\begin{figure}
 \begin{center}
 \includegraphics[scale=0.40]{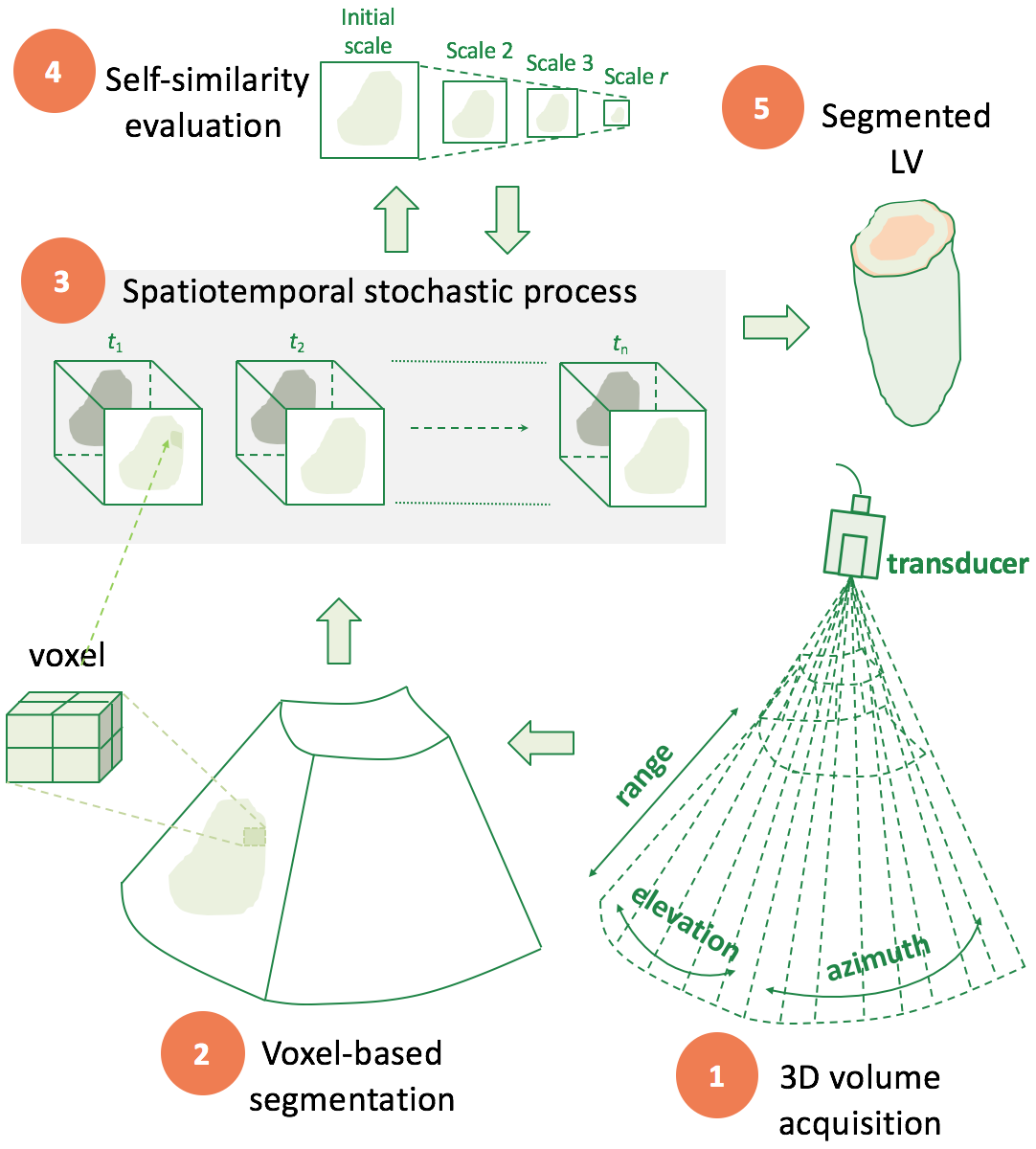}
 \end{center}
\caption{Illustrative example of different steps of multiscale parametric mapping based on fractional Brownian motion ($\tilde{B}_{H}$) for voxel-based segmentation.}
\label{fig:paramap}
\end{figure}

The fractal dimension (FD) of an $m$-dimensional fBm is related to the Hurst index \cite{alk17} by
\begin{equation}
    \label{FD}
    FD = m + 1 - H,
\end{equation}

where $m$ is the Euclidean dimension, or the number of independent variables, and $H$ quantifies the self-affinity of the process. Therefore, the closer $H$ is to one, the lower the FD, and the smoother the process becomes and vice versa. Hence, quantifying how smooth or regular the cardiac wall motion via fBm might assist in detecting abnormal left ventricular relaxation and increased stiffness related to myocardial ischemia and infarction \cite{thn06}. 

Finally the regional (or parametric) fractal dimension volume ($\bm{\mathcal{F}}$) is generated for each voxel in the time-sequence ultrasound images. In (\ref{eqn:fd_para_vol}), each voxel has its own localized Hurst index ($h$) after being subtracted from Euclidean dimension as in (\ref{FD}), and hence a fractal surface would represent the volume. The practical implementation is represented in Algorithm~\ref{alg:FD_map}.
\begin{equation}
\label{eqn:fd_para_vol}
\bm{\mathcal{F}}^{(\bm{i},\bm{r})} = \left(\begin{array} {cccc}
h^{i}_{11L} & h^{i}_{12L} & \cdots &  h^{i}_{1NL}		\\
h^{i}_{21L} & h^{i}_{22L} & \cdots &  h^{i}_{2NL}		\\
\vdots & \vdots & \ddots & \vdots 		\\
h^{i}_{M1L} & h^{i}_{M2L} & \cdots &  h^{i}_{MNL}		\\
\end{array}\right)
\end{equation}

The volume elements, namely $M$, $N$ and the coordinate index for the third dimension $L = \{l_1, l_2, \dots , l_m, \dots, l_z\}$ -- representing the slice position $l_m$ in the processed volume $V_z$, are defined for different volumes within the ultrasound time-sequence $\bm{i}$. The scaling factor $\bm{r} = 1, \ldots , j$ is the resolution limits of $\bm{\mathcal{F}}^{(\bm{i},\bm{r})}$, which represents the mean absolute intensity difference to center voxels based on local range dependence, cf. Fig.~\ref{fig:paramap_hurst}. The framework as conducted through $L$ voxel-by-voxel is illustrated in Fig.~\ref{fig:paramap}.

%\begin{figure}
% \begin{center}
% \includegraphics[scale=0.40]{figures/paramap}
% \end{center}
%\caption{\hl{Illustrative example of different steps of multiscale parametric mapping based on fractional Brownian motion} ($\tilde{B}_{H}$) \hl{for voxel-based segmentation.}}
%\label{fig:paramap}
%\end{figure}

\begin{algorithm} [t]
% \SetKwData{flag}{flag}
% \SetKwData{true}{true}
% \SetKwData{node}{node}
\KwIn{$\textbf{I}(x,y,z,i)$, $\textbf{I}$: 3D cardiac ultrasound sequences, in space $x, y, z$, and time $i$}
%\Parameter{Some parameter}
\While(\tcc*[h]{$i$:ultrasound time-sequence}){$i > 0$}
{
\ForEach{LV volume $V_1 \rightarrow V_z$}{
\textbf{Step 1:} 3D shape modeling using surface parameterization\\
\ForAll{voxels $v$ in cuboid lattice $l$}{
\textbf{Step 2:} Initialize self-similarity stochastic process\\
\ForAll{voxel pair distances $\Delta r$ in $l$}{
Compute mean absolute difference $\Delta v$ of each voxel pair $p_i, q_i$;\\
Normalize and take the logarithm $\Delta \hat{v} = log\left(\Delta v^i_{mnr}/\left\|\Delta v^i_{mnr}\right\|\right)$; \tcc*[h]{$m$, $n$ and $r$ are size of voxel $v_i$ at certain scale $j$}\\
Normalize voxel pairs distances $\Delta\hat{r}$; \tcc*[h]{s.t. $\Delta\hat{r} = \sqrt{\sum_{i=1}^{n}\left(q_i-p_i\right)^2}$} \\
Perform least square linear regression as: $S_{rr} = \sum^{j}_{s=1} \Delta \hat{r}^2_s - {\left(\sum^{j}_{s=1}\Delta\hat{r}_s\right)^2}/{j}$,\\
$S_{rv} = \sum^{j}_{s=1}\sum^{j}_{k=1}\Delta \hat{r}_s \hat{v}_k - {\left(\sum^{j}_{s=1} \Delta\hat{r}_s\right)\left(\sum^{j}_{k=1}\Delta\hat{							v}_k\right)}/{j}$;\\
\textbf{Step 3:} Hurst index \textit{\textbf{H}} matrix\\ 
Estimate localized Hurst indices ($h$): $ h = \left(S_{rv}/S_{rr}\right)$; \tcc*[h]{local linear regression}\\
Generate fractal map:  $\mathcal{F}^{h}[i,r] \leftarrow 3 - H$; \tcc*[h]{for time-sequence $i$ and scaling factor $r$}
}
}
Estimate surface heterogeneity from $\bm{\mathcal{F}}^{h}[i,r]$;\\
Construct feature vector for Volume $V_k$: $\bm{\lambda}_k = \{f^1_{i,r}, \dots,f^k_{i,r}, \dots, f^z_{i,r}\}$;
}
$i \leftarrow i-1$
}
\Return $\{\bm{\mathcal{F}}^h, \bm{\lambda}_k\}$\;
\KwOut{Fractal parametric voxel map and features vector $\{\bm{\mathcal{F}}^h, \bm{\lambda}_k\}$}    
\caption{Fractal parametric voxel map construction}\label{alg:FD_map}
\end{algorithm}

\begin{figure*}
 \begin{center}
 \includegraphics[scale=0.35]{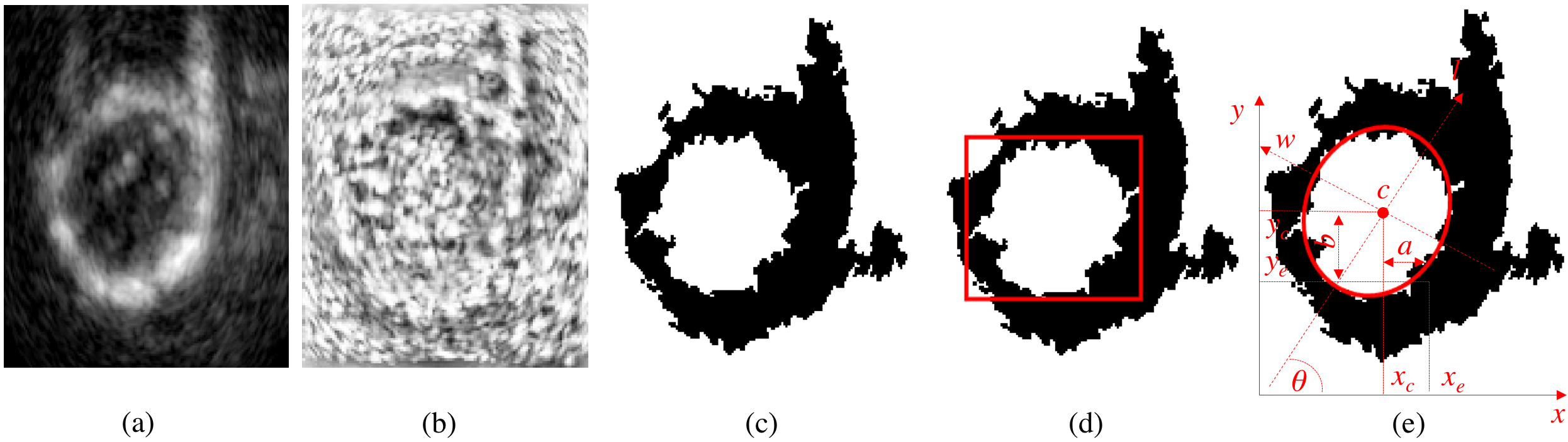}
 \end{center}
\caption{Endocardium segmentation: (a) B-mode 2D ultrasound image, (b) corresponding fractal parametric voxel map (c) fBm segmented binary image, (d) moment-based endocardium region based on enclosing rectangle and (e) ellipsoidal approximation, respectively.}
\label{fig:endo_seg}
\end{figure*}

\subsection{Classification}
Patches of $6.3\times8.1\times5.4$ mm$^3$ -- representing myocardium vs blood pool -- were selected randomly from a single ultrasound volume transformed to $\bm{\mathcal{F}}$, and used for training data. Such that, the training set, $\bm{S_T}$, is composed of a set of $N$ patches for which the feature vector $\bm{V}$, and the classification result ($C_1$ or $C_2$: myocardium or blood pool) are known $\bm{S_T} = \{\bm{V}^{(\bm{n})},\bm{C}_{k}^{(\bm{n})} |\bm{n} = 1, \dots,N; k \in \{1,2\} \}$. The Hurst index is a useful statistical method for inferring the properties of a time series without making assumptions about stationarity. It is most useful when used in conjunction with other techniques. Thus, features representing the FD mean and variance values over $\bm{\mathcal{F}}$, lacunarity ($\mathcal{L}$) -- which defines the sparsity of the fractal pattern in terms of the ratio of the variance over the mean of $\bm{\mathcal{F}}$ as in (\ref{eqn:lac}) \cite{alk17}:
\begin{equation}
\label{eqn:lac}
    \mathcal{L} = \frac{\frac{1}{MNK}\sum_{x=1}^M\sum_{y=1}^N\sum_{z=1}^K\bm{\mathcal{F}}(\textbf{x})^2}{\left(\frac{1}{MNK}\sum_{x=1}^M\sum_{y=1}^N\sum_{z=1}^K\bm{\mathcal{F}}(\textbf{x})\right)^2},
\end{equation}
and higher order statistics, namely, skewness (asymmetry of the probability distribution) and kurtosis (fourth standardized moment) are defined over localized histograms derived from $\bm{\mathcal{F}}$ and normalized to form a 5--D vector in feature space $\bm{\lambda}=(f_1,f_2,\dots,f_5)$. Then in the classification procedure, one of the classes $C_1$ (myocardium) or $C_2$ (blood pool) would be assigned to each candidate voxel when its representation is known. Since our primary concern is to demonstrate the robustness of the new multiscale parametric mapping based on fBm, we describe experiments conducted to compare the discriminative power -- which relies naturally on the spatio-temporal dependencies of the local descriptors -- for classification. A Bayesian classifier, being a simple probabilistic and commonly used machine learning benchmarking method, which performs well even with possible presence of dependent attributes \cite{dmg97} was selected. This particularly suits fractal patterns in ultrasound images, where not all locally estimated features are conditionally independent given the class.%A Bayesian classifier was used for classification, although SVM or random forests would have also served the purpose. 

The sample training patches were collected from manually labeled LV (performed by 2 experts to avoid errors due to ambiguous interpretation of structures) cavity--blood pool and myocardium voxels in the synthetic and real training images. Specifically, around $\sim45200$ and $\sim35000$ voxel samples referring to 62 and 48 voxel patches for the synthetic and animal dataset, respectively, were fairly divided into myocardium and blood pool voxel patches and used for training. A total of 62 and 48 sample voxel patches were used for myocardium training, for synthetic and animal dataset. Similarly applies for the blood pool training. The training voxels represent 0.41\% and 0.29\% of the total number of synthetic and animal dataset test image voxels, respectively. To reduce the risk of introducing errors, and thus noise in the classification stage, training samples were carefully selected to cover all possible regions related to blood pool, endocardium, epicardium, and myocardium. It is worth noting that the LV segmentation was performed voxel-by-voxel in 3D by employing a multiresolution fBm-based kernel operating as a sliding-window (i.e. cuboid $v_{m \times n \times r}$) in a 3D volume sequence ($V^{(i)}_{M\times N\times L}$); revisit Algorithm~\ref{alg:FD_map}. This would give smoother segmentation in homogeneous tissue regions while preserving fine details in heterogeneous regions, viz. high changes in tissue specular reflections.

\subsection{Ellipsoidal model assignment to fBm maps}
\label{method_mont}

For an improved refinement of the LV contour boundaries, each of the fBm segmented regions are fitted by means of image moments to an elliptical model. Moments of images can provide efficient local descriptors and have been used extensively in image analysis applications \cite{kls07,sit14}. Their main advantage is their ability to provide invariant measures of shape, which can better characterize the heterogeneity of the LV wall surface.

Image moments can be defined as weighted averages of voxel intensities. For the time-sequence 3D ultrasound images, at time $t$ and depth $z$, the raw $(p,q)$-moment $\bm{m}_{p,q}$ for an fBm segmented object $\bm{G}$ (LV for our case) is given by:
\begin{equation}
    \label{}
    \bm{m}_{pq} = \sum_{x=1}^{M}\sum_{y=1}^{N}x^{p}y^{q}f(x,y),
\end{equation}
\noindent where $M$ and $N$ are the size of a 2D-image slice of a 3D object $\bm{G}$ and $f(x,y)$ are the labeled binary output of the calculated FD values.

The first-order moments $m_{1,0}$ and $m_{0,1}$, when normalized by $m_{0,0}$ give the coordinates of the binary object -- $x_c$ and $y_c$ of the endocardium or epicardium. Accordingly, second-order moments describe the ``distribution of mass'' with respect to the coordinate axes and define the orientation $\theta$ of $\bm{G}$. In order to extract the parameters of the equivalent ellipse from the second-order moments $m_{2,0}$, $m_{1,1}$, and $m_{0,2}$, the central moments can be defined as

\begin{equation}
    \label{eqn:momnt}
    \bm{\mu}_{pq} = \sum_{p,q \in \bm{G}}^{N_g}(\bm{x} - x_c)^p(\bm{y} - y_c)^q,
\end{equation}
such that
\begin{gather}
    \label{eqn:moment_2d}
    \mu_{2,0} = \frac{m_{2,0}}{m_{0,0}} - x_{c}^2, \quad \mu_{1,1} = 2(\frac{m_{1,1}}{m_{0,0}} - x_{c}y_{c}), \nonumber\\ \mu_{0,2} = \frac{m_{0,2}}{m_{0,0}} - y_{c}^2,
\end{gather}
\noindent where $x_c$ and $y_c$ are the coordinates of the centroid $c$ of $\bm{G}$ having a size of $N_g$. These moments are invariants to translation. Then the covariance matrix of the binary object would be:\\
$$cov(\bm{G}) = \left( 
\begin{array}{cc}
    \mu_{2,0} & \mu_{1,1} \\
    \mu_{1,1} & \mu_{0,2} \\
\end{array} \right)$$

\noindent and the eigen vectors of this covariance matrix correspond to the major and minor axes of the equivalent ellipse.

From (\ref{eqn:momnt}) the summation extends over all elements in $\bm{G}$, such that $\mu_{0,0}$ represents the area of the pattern, i.e., the number of white pixels in the middle of Fig~.\ref{fig:endo_seg}(c). The coordinates of the centroid $c = (x_c,y_c)$ can be calculated combining $\mu_{0,0}$ with the image moments of the first degree $\mu_{0,1}$ and $\mu_{1,0}$. The binary image of the equivalent rectangle\footnote{The farthest pixel from the centroid of $\bm{G}$ is specified as the upper-left corner of the rectangle and perpendicular lines are projected in each dimension. Then the smallest bounding box enclosing $\bm{G}$ in $f(x,y)$ is found iteratively.} in Fig~.\ref{fig:endo_seg}(d) has the same zeroth, first and second moments. Using the moments of the second degree $\mu_{1,1}$, $\mu_{0,2}$ and $\mu_{2,0}$, the final formulae giving the major axis orientation $\theta$ and the respective major and minor axis lengths $l$ and $w$, are calculated as follows:
\begin{gather}
    \theta = \frac{1}{2}tan^{-1}\bigg(\frac{2\mu_{1,1}}{\mu_{2,0}-\mu_{0,2}}\bigg), \\
    l = \sqrt{6\Big(\mu_{2,0}+\mu_{0,2}+\sqrt{4\mu_{1,1}^{2}+(\mu_{2,0}-\mu_{0,2})^2}\Big)}, \\
    w = \sqrt{6\Big(\mu_{2,0}+\mu_{0,2}-\sqrt{4\mu_{1,1}^{2}+(\mu_{2,0}-\mu_{0,2})^2}\Big)}\label{ellipse_param}.
\end{gather}

Having these parameters we can infer the equivalent ellipse, where $c$ was first identified on the fBm segmented LV base image, and used afterwards for alignment -- throughout the rest of image sequence -- towards the apex. Fig~.\ref{fig:endo_seg} illustrates the above concepts, where $a = \mu_{2,0}$ and $b = \mu_{1,1}$. In this paper, we employ enclosing elements with dimensions $(2w, 2l)$.

%The relative difference in magnitude of the eigenvalues $w$  and $l$ are thus an indication of the eccentricity of the image, or how elongated it is. The eccentricity is expected to vary as the LV shape change is $\sqrt{1-(w/l)}$, where an ellipse whose eccentricity is $0$ is actually a circle. 

\section{Experimental Results}
\label{results}
\subsection{Experimental Setup}

The framework robustness was validated on five different synthetic 3D ultrasound sequences simulating normal and ischemic LV conditions, and furthermore on real 3D cardiac ultrasound sequences acquired from two canine subjects under resting-state and stress-state conditions.

\subsubsection{Synthetic Data} 
Validation was done on 3D ultrasound time-sequences from the KU Leuven synthetic dataset \cite{asd15}, representing one normal and four ischemic cases. Therein, different cases related to LV healthy and pathological conditions were generated. In the mechanical simulations, normal contractility and stiffness values -- according to the standard American Heart Association \cite{crq02} -- were assigned for the LV normal geometry. The mechanical parameters were tuned in order to match ejection fraction measured on the corresponding template acquisition \cite{asd15}. From the healthy geometry, four ischemic simulations were generated by altering contractility and stiffness in diseased segments. The four ischemic simulations corresponded to: a distal and proximal occlusion of the Left Anterior Descending artery (LADdist and LADprox respectively); occlusions of Right Coronary Artery (RCA) and Left Circumflex (LCX).

\begin{figure}[th]
 \begin{center}
 \includegraphics[scale=0.50]{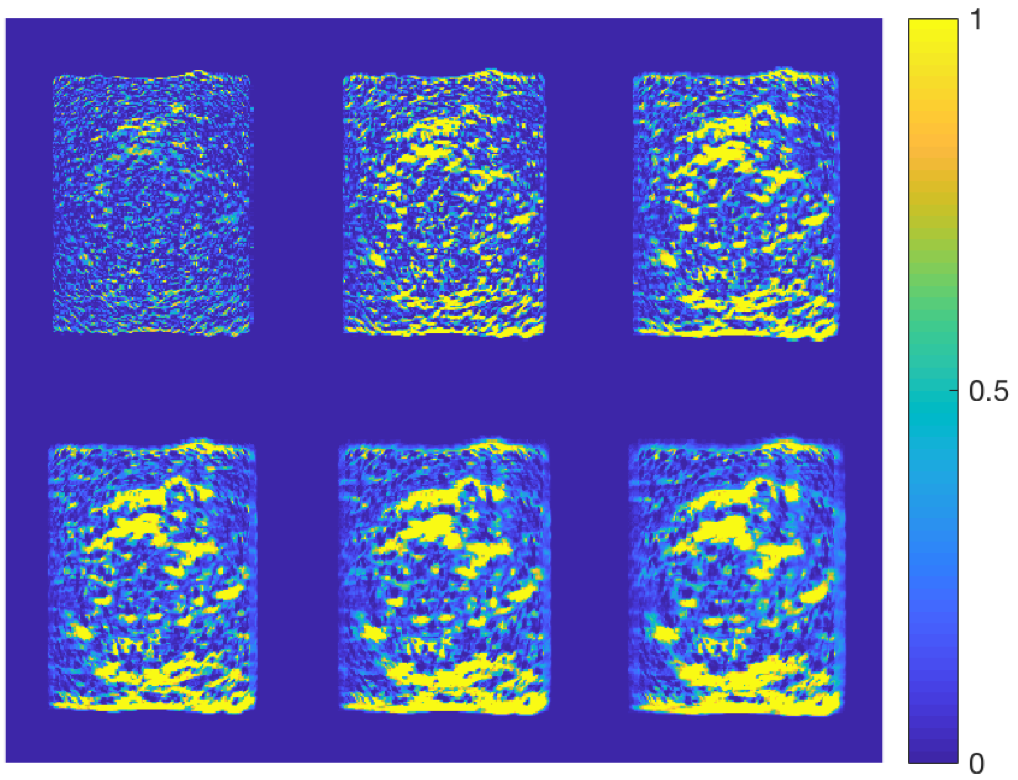}
 \end{center}
\caption{Measuring the discrepancy by means of residual sum of squares -- bright regions indicate higher error --between a $224 \times 176$ sample cardiac ultrasound sequence image with a voxel size of $0.7 \times 0.9$ mm$^2$ (shown in Fig.~3(a)) and fBm estimation model with different localized range dependence ($2.1 \times 2.7$ mm$^2$, $3.5 \times 4.5$ mm$^2$, $4.9 \times 6.3$ mm$^2$, $6.3 \times 8.1$ mm$^2$, $7.7 \times 9.9$ mm$^2$, $9.1 \times 11.7$ mm$^2$, respectively).}
\label{fig:fBm_RSS}
\end{figure}

Each image in the dataset had $224\times176\times208$ voxels of size $0.7\times0.9\times0.6$ mm$^3$. On average there were 35 images per sequence. For each sequence, ground truth motion trajectories were provided at 2250 mesh points. The endocardium and epicardium meshes were then converted into segmentation masks and used for benchmarking. 

\subsubsection{Animal Data}
3D ultrasound sequence images were acquired from two acute canine studies (open chested) following a severe occlusion of the left anterior descending coronary artery. A Philips iE33 ultrasound imaging system (Philips Health Care, Andover, MA) with a X7-2 probe at 4.4 MHz suspended in a water bath over the heart was used for image acquisition. Acquisition time points included baseline and one hour and 6 weeks after surgical occlusion of the left anterior descending coronary artery. Images typically had $400\times140\times120$ voxels of $0.25\times0.85\times0.85$ mm$^3$ with an average of 23 temporal frames. All experiments were conducted in compliance with the Institutional Animal Care and Use Committee policies.

\subsection{Evaluating Model Goodness-of-Fit}
The residual sum of squares (RSS) was used to measure the discrepancy in the estimated $H$ indices of the parametric volume maps ($\bm{\mathcal{F}}$). A low value indicates the model has a smaller random error component. Accordingly, the surface of $\bm{\mathcal{F}}$ was initially estimated using different fBm localized range dependence by adjusting the range of the scaling factor $\bm{r}$ in (\ref{eqn:fd_para_vol}). Higher RSS was encountered in longer range dependence -- i.e. higher resolutions for $\bm{r}$; revisit Fig.~\ref{fig:paramap}-- as the introduced error may get accumulated due to echo artifacts, and hence the $H$ estimation becomes unstable. % as shown in Fig.~\ref{fig:fBm_Drange}. 
In order to graphically represent the method estimation discrepancy, as this can give a better understanding of the relationship between the fBm model and the surface of the speckle pattern, Fig.~\ref{fig:fBm_RSS} shows where errors most likely occur when locally computing $H$, and hence the estimation of $\bm{\mathcal{F}}$. Selecting the best fBm dependence range can avoid unnecessary computational time and give better accuracy.

\begin{figure}[th]
 \begin{center}
 \includegraphics[scale=0.35]{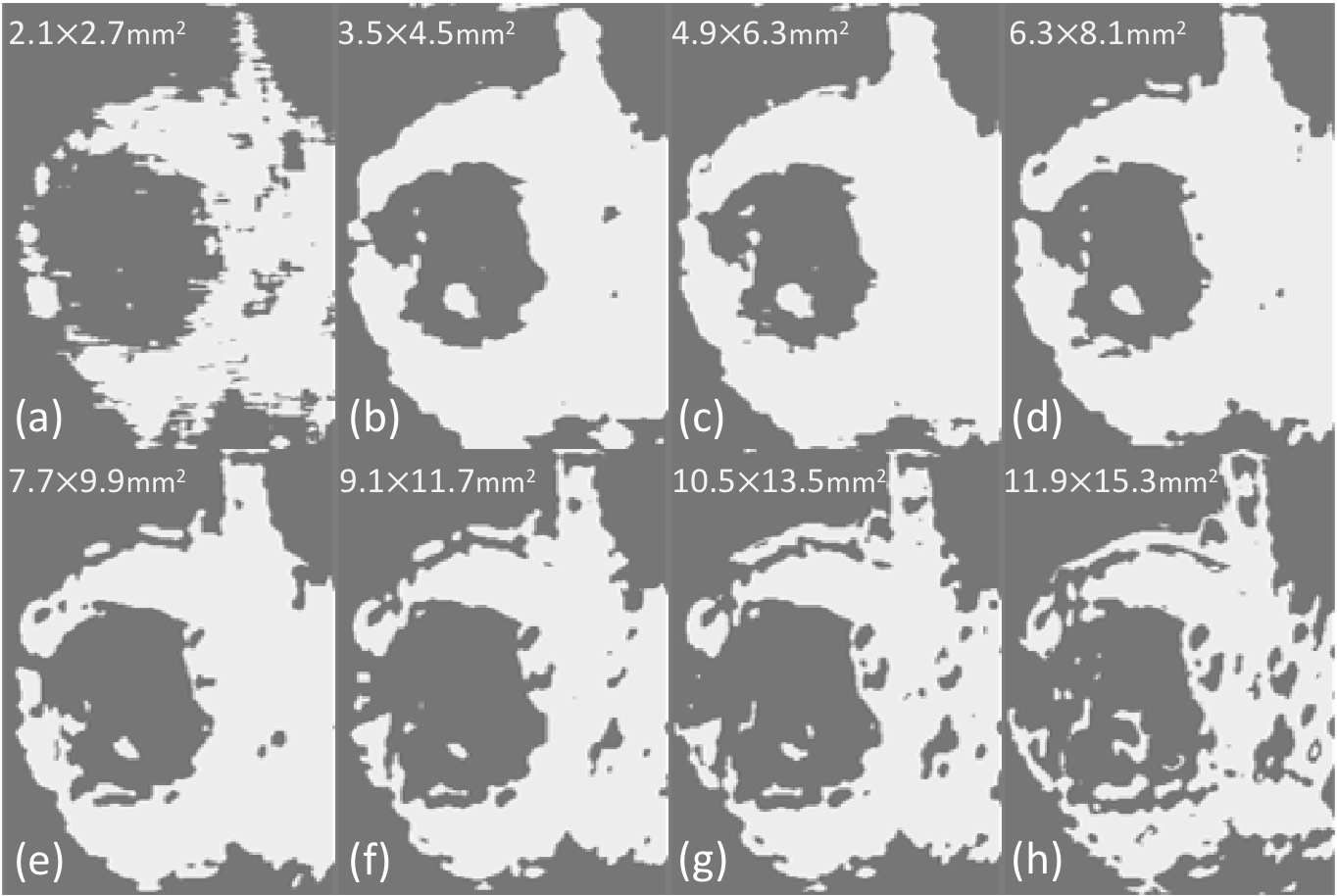}
 \end{center}
\caption{Left ventricle endocardium and epicardium initial segmentation result for a sample cardiac ultrasound sequence image by varying the localized fBm dependence range. Best delineation of the endocardium versus epicardium boundary is shown in (c).}
\label{fig:fBm_size}
\end{figure}

\subsection{Patch size vs localized fBm}
\label{psz_locfBm}

%The patch size refers to how large is the neighborhood area when locally estimating the fBm for each voxel. The trade-off between patch size and texture pattern in the ultrasound image is a well-known challenge, and a balance between the two parameters can improve the segmentation quality. A varying patch size is shown in Fig.~\ref{fig:wsize} to better localize the various possible structures encountered in image texture. The best patch size is the one where the blood-pool appears as an island in the myocardium, and is intact with no disconnected regions, which is Fig.~\ref{fig:wsize}(g) having a size of $10.5 \times 13.5$ mm$^2$.

\begin{figure}[!ht]
 \begin{center}
 \includegraphics[scale=0.30]{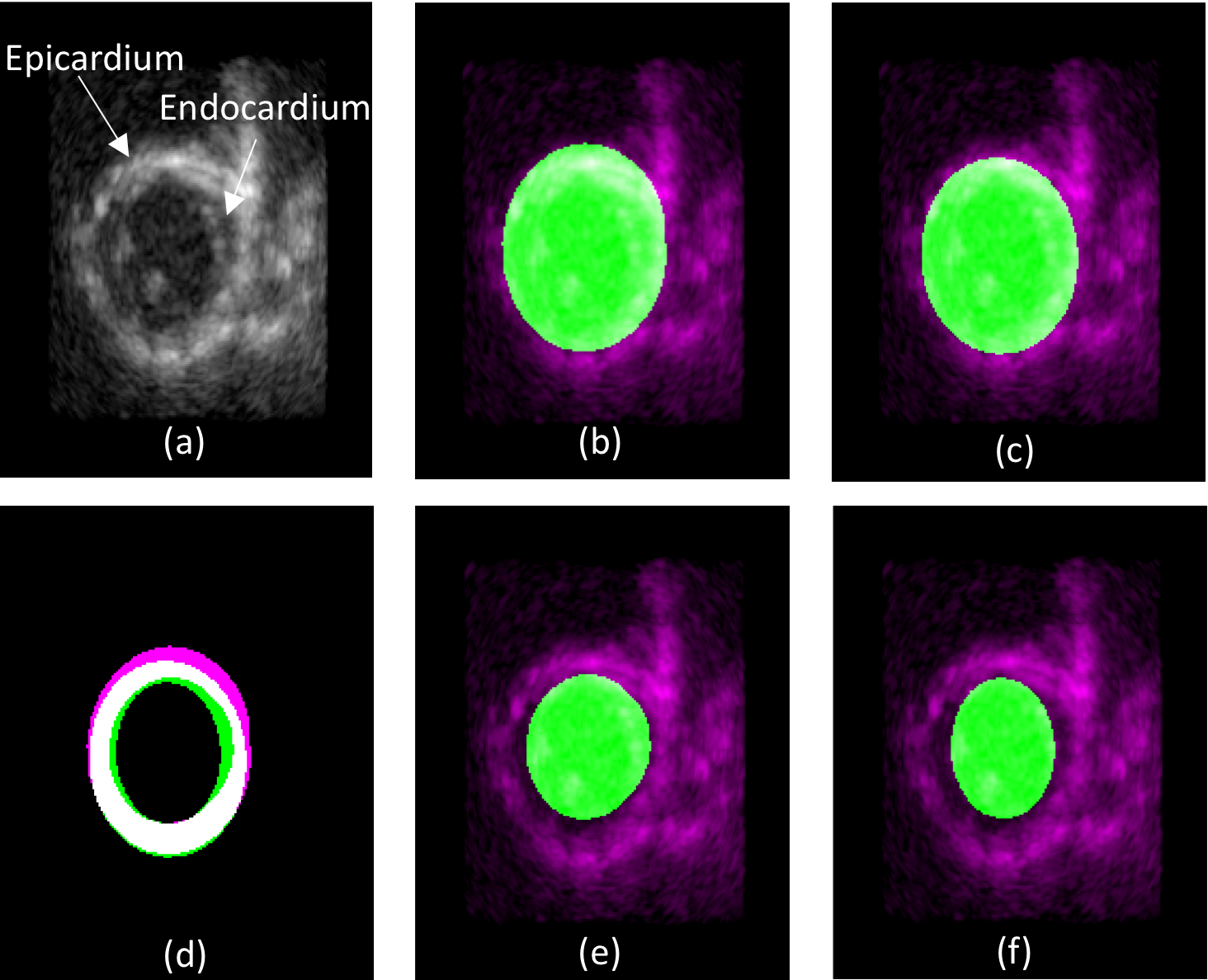}
 \end{center}
\caption{Segmentation of the left ventricle of the heart from (a) 2D time sequence ultrasound image demonstrating in (b) and (c) epicardium ground-truth vs automatic fBm segmentation, (e) and (f) endocardium ground-truth vs automatic fBm segmentation, respectively. A joint comparison between ground-truth vs automatic left ventricle fBm segmentation is shown in (d), where white color means perfect match, while magenta and green refer to dissimilarity of the ground truth vs automatic fBm segmentations, respectively. (For qualitative interpretation of segmentation quality, figures (b)-(f) best viewed in color)}
\label{fig:LV_seg_2D}
\end{figure}

The local estimation of the fBm process by indicating how far the resolution of $\bm{r}$ can be deeply probed is an important factor for improving segmentation quality. Unlike mathematical fractals where they tend to scale infinitely, the localized fBm dependence range -- calculating how deeply the resolution limits of $\bm{r}$ can be probed -- is essential for an improved segmentation of LV boundaries. Fig.~\ref{fig:fBm_size} empirically shows how the $H$ values vary at different fBm local dependence ranges, with an indication of the best possible range. To insure consistency, the dependence range has been varied in Fig.~\ref{fig:fBm_size} in a similar ratio to how the patch size was changed. The best localized range that better delineates the blood pool from the epicardium with the least segmentation errors is Fig.~\ref{fig:fBm_size}(c). An example of a LV base segmentation after undergoing fBm stochastic modelling and ellipse fitting for boundary refinement is illustrated in Fig.~\ref{fig:LV_seg_2D}. %\footnotetext{Physical interpretation of FD is straightforward: the closer FD to 2 the process becomes smoother. Conversely, the closer FD to 3, the process is closer to a random walk, and hence more heterogeneous.}

%\begin{figure}[th]
% \begin{center}
% \includegraphics[scale=0.35]{figures/wsize}
% \end{center}
%\caption{Effect of varying patch-size while fixing the localized fBm range dependence on the endocardium and epicardium segmentation. Patch-size increase by a step-size of $1.4 \times 1.8$ mm$^2$ from (a) $2.1 \times 2.7$ mm$^2$ to (h) $11.9 \times 15.3$ mm$^2$.}
%\label{fig:wsize}
%\end{figure}
%\clearpage

\subsection{Quantitative Evaluation}
\begin{figure}
 \begin{center}
 \includegraphics[scale=0.37]{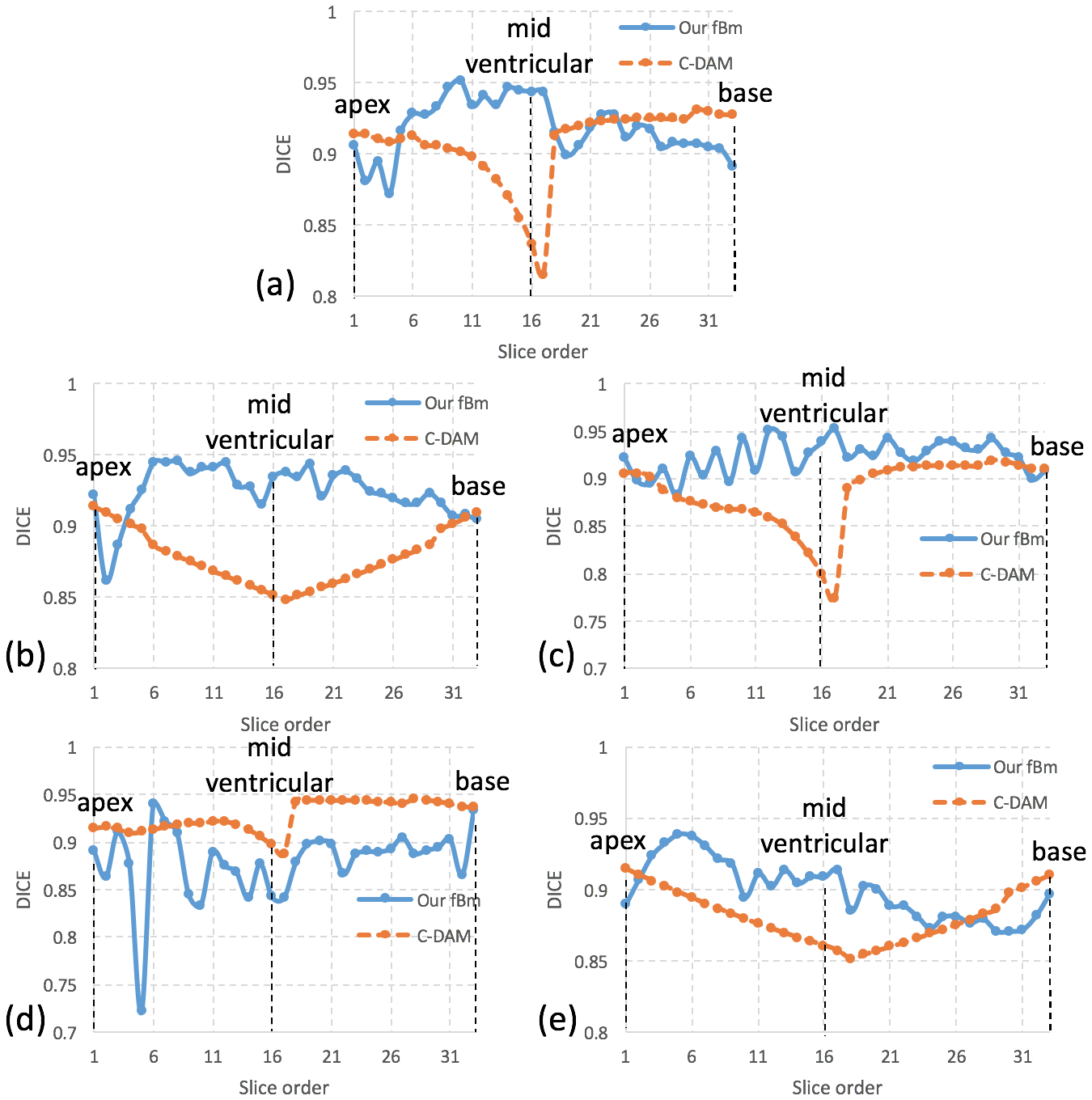}
 \end{center}
\caption{Comparison of slice-by-slice dice coefficient segmentation quality for fBm and C-DAM \cite{hng14} in (a) normal, (b) RCA, (c) LCX, (d) LADdist, and (e) LADprox left ventricle conditions.}
\label{fig:slice}
\end{figure}

The proposed fBm segmentation method was benchmarked using state-of-the-art method in \cite{hng14}, which is based on employing spatio-temporal coherence under a dynamic appearance model, abbreviated herein as C-DAM. The evaluation of the segmentation quality was performed using three different segmentation measures, which are: Dice coefficient (DC), Hausdorff distance (HD), and mean absolute distance (MAD), and computed for both end-diastolic and -systolic and averaged over all cases, see Table \ref{table:endo_class} and \ref{table:epi_class}. Fig.~\ref{fig:slice} shows a slice-by-slice LV epicardium segmentation quality evaluation corresponding to the DICE metric in Table \ref{table:epi_class}. Also the results of the resting/stress state of the canine dataset showed an improvement of $2.4\%$, $0.19$mm, $0.37$mm and $1.1\%$, $0.08$mm, $0.17$mm for the DC, HD, and MAD segmentation quality metrics of the epicardium and endocardium borders for the fBm-based as compared to the C-DAM method, respectively.

For clinical relevance, as  numerical measures tend to be more narrowly focused on a particular aspect of the data and often try to compress information into a single number, the LV volume segmentation quality is represented as well graphically in Fig.~\ref{fig:LV_seg_3D} and Fig.~\ref{fig:LV_vol_cn_mid_slice} for the synthetic and animal dataset, respectively. Results show that the method in \cite{hng14} tends to over-segment the epicardium boarder by extending into the surrounding area of myocardium (i.e. LV appearing dilated as compared to ground truth), and the epicardium surface is more irregular as compared to the proposed fBm-based segmentation method, cf. Fig.~\ref{fig:LV_seg_3D}(f) and (k). Also, the mid-ventricular cavity shows improved delineation in the complex canine dataset using the fBm-based segmentation method, cf. Fig.~\ref{fig:LV_vol_cn_mid_slice}(g)-(i). Furthermore, the absolute difference is measured between the different voxel patches in search for self-similarity properties, i.e. invariance under a suitable change of scale. It describes the different statistical distribution variations found in the LV ventricular cavity and myocardium. To this end, the estimated fBm local range dependence serves in determining how far the image resolution can be deeply probed. Terminating the search at the optimized scaling factor $r$ contributes in improving segmentation of the LV boundaries, and further saves unnecessary computational time. The implemented fBm algorithm in this paper had a computational complexity of $O(Nlog(N))$, with running time nearly similar for both methods, i.e. around 1 minute per frame. However, the proposed fBm segmentation method is fully automatic and does not rely on training data.

%%%%%%%%%%%%%%%% Endocardium Segmentation Quality %%%%%%%%%
\begin{table}[!t]
\centering
\caption{Endocardium segmentation for normal and ischemic cardiomyopathy}
\label{table:endo_class}
\begin{tabular}{@{}lllll@{}}
\toprule
\multirow{2}{*}{\begin{tabular}[c]{@{}l@{}}Left Ventricle \\ Condition\end{tabular}} & \multirow{2}{*}{Method} & \multicolumn{3}{c}{Segmentation Evaluation Metric} \\ \cmidrule(lr){3-5} 
                                          &                         & DICE (\%)        & HD (mm)        & MAD (mm)       \\ \cmidrule(lr){1-5}
\multirow{2}{*}{\begin{tabular}[c]{@{}l@{}}Normal \\ {}\end{tabular}}                    & C-DAM                   & $83.87 \pm 5.79$   & $2.66 \pm 0.16$  & $0.47 \pm 0.16$  \\
                                          & Our fBm                 & $\textbf{87.08} \pm 1.63$   & $\textbf{2.64} \pm 0.13$  & $\textbf{0.44} \pm 0.07$  \\ \cmidrule(lr){2-5}
\multirow{2}{*}{\begin{tabular}[c]{@{}l@{}}RCA \\ {}\end{tabular}}           & C-DAM                   & $84.87 \pm 4.43$   & $2.74 \pm 0.22$  & $0.47 \pm 0.12$  \\
                                          & Our fBm                 & $\textbf{88.15} \pm 1.87$   & $\textbf{2.65} \pm 0.17$  & $\textbf{0.42} \pm 0.08$  \\\cmidrule(lr){2-5}
\multirow{2}{*}{\begin{tabular}[c]{@{}l@{}}LCX \\ {}\end{tabular}}           & C-DAM                   & $82.29 \pm 4.00$   & $2.70 \pm 0.14$  & $0.57 \pm 0.12$  \\
                                          & Our fBm                 & $\textbf{87.43} \pm 1.40$   & $\textbf{2.58} \pm 0.10$  & $\textbf{0.46} \pm 0.07$  \\\cmidrule(lr){2-5}
\multirow{2}{*}{\begin{tabular}[c]{@{}l@{}}LADdist \\ {}\end{tabular}}       & C-DAM                   & $\textbf{84.34} \pm 4.62$   & $\textbf{2.72} \pm 0.18$  & $\textbf{0.53} \pm 0.15$  \\
                                          & Our fBm                 & $83.66 \pm 3.92$   & $2.83 \pm 0.16$  & $0.56 \pm 0.12$  \\\cmidrule(lr){2-5}
\multirow{2}{*}{\begin{tabular}[c]{@{}l@{}}LADprox \\ {}\end{tabular}}       & C-DAM                   & $\textbf{85.82} \pm 3.04$   & $\textbf{2.76} \pm 0.15$  & $\textbf{0.44} \pm 0.10$  \\
                                          & Our fBm                 & $85.18 \pm 2.24$   & $\textbf{2.76} \pm 0.15$  & $0.48 \pm 0.10$  \\ \bottomrule 
\end{tabular}
\end{table}

%\begin{figure}[ht]
% \begin{center}
% \includegraphics[scale=0.30]{figures/LV_seg_2D}
% \end{center}
%\caption{Segmentation of the left ventricle of the heart from (a) 2D time sequence ultrasound image demonstrating in (b) and (c) epicardium ground-truth vs automatic fBm segmentation, (e) and (f) endocardium ground-truth vs automatic fBm segmentation, respectively. A joint comparison between ground-truth vs automatic left ventricle fBm segmentation is shown in (d), where white color means perfect match, while magenta and green refer to dissimilarity of the ground truth vs automatic fBm segmentations, respectively. (For qualitative interpretation of segmentation quality, figures (b)-(f) best viewed in color)}
%\label{fig:LV_seg_2D}
%\end{figure}

%%%%%%%%%%%%% Epicardium Segmentation Quality %%%%%%%%%%%%%
\begin{table}[!t]
\centering
\caption{Epicardium segmentation for normal and ischemic cardiomyopathy}
\label{table:epi_class}
\begin{tabular}{@{}lllll@{}}
\toprule
\multirow{2}{*}{\begin{tabular}[c]{@{}l@{}}Left Ventricle \\ Condition\end{tabular}} & \multirow{2}{*}{Method} & \multicolumn{3}{l}{Segmentation Evaluation Metric}   \\ \cmidrule(l){3-5} 
                                                                                     &                         & DICE (\%)        & HD (mm)         & MAD (mm)        \\ \cmidrule(lr){1-5}
\multirow{2}{*}{\begin{tabular}[c]{@{}l@{}}Normal \\ {}\end{tabular}}                                                              & C-DAM                   & $90.68 \pm 2.34$ & $2.89 \pm 0.37$ & $0.73 \pm 0.24$ \\
                                                                                     & Our fBm                 & $\textbf{91.88} \pm 1.88$ & $\textbf{2.72} \pm 0.21$ & $\textbf{0.66} \pm 0.27$ \\\cmidrule(lr){2-5}
\multirow{2}{*}{\begin{tabular}[c]{@{}l@{}}RCA \\ {}\end{tabular}}                                                      & C-DAM                   & $87.78 \pm 2.19$ & $2.91 \pm 0.22$ & $0.99 \pm 0.14$ \\
                                                                                     & Our fBm                 & $\textbf{92.47} \pm 1.70$ & $\textbf{2.84} \pm 0.18$ & $\textbf{0.70} \pm 0.24$ \\\cmidrule(lr){2-5}
\multirow{2}{*}{\begin{tabular}[c]{@{}l@{}}LCX \\ {}\end{tabular}}                                                      & C-DAM                   & $88.50 \pm 3.10$ & $2.93 \pm 0.48$ & $1.11 \pm 0.40$ \\
                                                                                     & Our fBm                 & $\textbf{92.32} \pm 1.68$ & $\textbf{2.64} \pm 0.15$ & $\textbf{0.74} \pm 0.24$ \\\cmidrule(lr){2-5}
\multirow{2}{*}{\begin{tabular}[c]{@{}l@{}}LADdist \\ {}\end{tabular}}                                                  & C-DAM                   & $\textbf{92.72} \pm 1.45$ & $\textbf{2.80} \pm 0.31$ & $\textbf{0.61} \pm 0.21$ \\
                                                                                     & Our fBm                 & $87.98 \pm 3.67$ & $2.92 \pm 0.19$ & $1.02 \pm 0.31$ \\\cmidrule(lr){2-5}
\multirow{2}{*}{\begin{tabular}[c]{@{}l@{}}LADprox \\ {}\end{tabular}}                                                  & C-DAM                   & $88.06 \pm 1.92$ & $2.88 \pm 0.18$ & $0.95 \pm 0.06$ \\
                                                                                     & Our fBm                 & $\textbf{89.95} \pm 2.21$ & $\textbf{2.80} \pm 0.28$ & $\textbf{0.81} \pm 0.24$ \\ \bottomrule 
\end{tabular}
\end{table}

\section{Discussion}
\label{discuss}

The fBm process is a useful stochastic method for inferring the properties of a time series without making assumptions about stationarity. It relates to how strong the autocorrelations of the spatio-temporal coherence, and the rate at which these decrease as the lag between pairs of scatterer patterns increases. Although it might seem that searching for self-similarity properties by probing for higher resolutions, i.e. selecting higher values for $\bm{r}$ in (\ref{eqn:fd_para_vol}), could improve the segmentation quality, results show that reliable estimate of $H$, and hence the $\bm{\mathcal{F}}$, is valid only at a certain cutoff scale where there can be no more details. That is, the echo patterns being approximate rather than deterministic, as the characteristics of the pattern tends to scale in a statistical fashion. The self-similarity property in this sense means invariance in distribution under a suitable change of scale $\bm{r}$. Therefor the local linear regression relation for estimating $H$ (revisit Algorithm~\ref{alg:FD_map}) becomes non-linear at higher values of $\bm{r}$ -- due to exceeding the actual resolution of the ultrasound image itself \cite{alk08}, and eventually resulting in error accumulation. On the other hand, complementing the self-similarity property of the fBm process with the lacunarity measure can give a better characterization of the LV shape heterogeneity. Lacunarity analysis is a technique introduced to deal with fractal objects of the same dimension with different textural appearances \cite{alk09}. The lacunarity parameter describes the local complexity of the speckle pattern, on different scales, based on the spatial distribution of gaps of a specific size. Namely, it is a measures of the sparsity of the fBm process, providing additional information on the how irregularity fills the space.

%\begin{figure}
% \begin{center}
% \includegraphics[scale=0.29]{figures/LV_seg_3D}
% \end{center}
%\caption{3D segmentation volumes of the left ventricle for 3D echocardiographic sequences illustrating [upper-row] ground truth for normal, ischemic-RCA, ischemic-LCX, ischemic-LADdist, and ischemic-LADprox; [mid-row] C-DAM method in \cite{hng14}; [bottom-row] proposed fully-automated fBm approach. (Video examples can be referred to in supplementary materials)}
%\label{fig:LV_seg_3D}
%\end{figure}

\begin{figure} [!th]
 \begin{center}
 \includegraphics[scale=0.29]{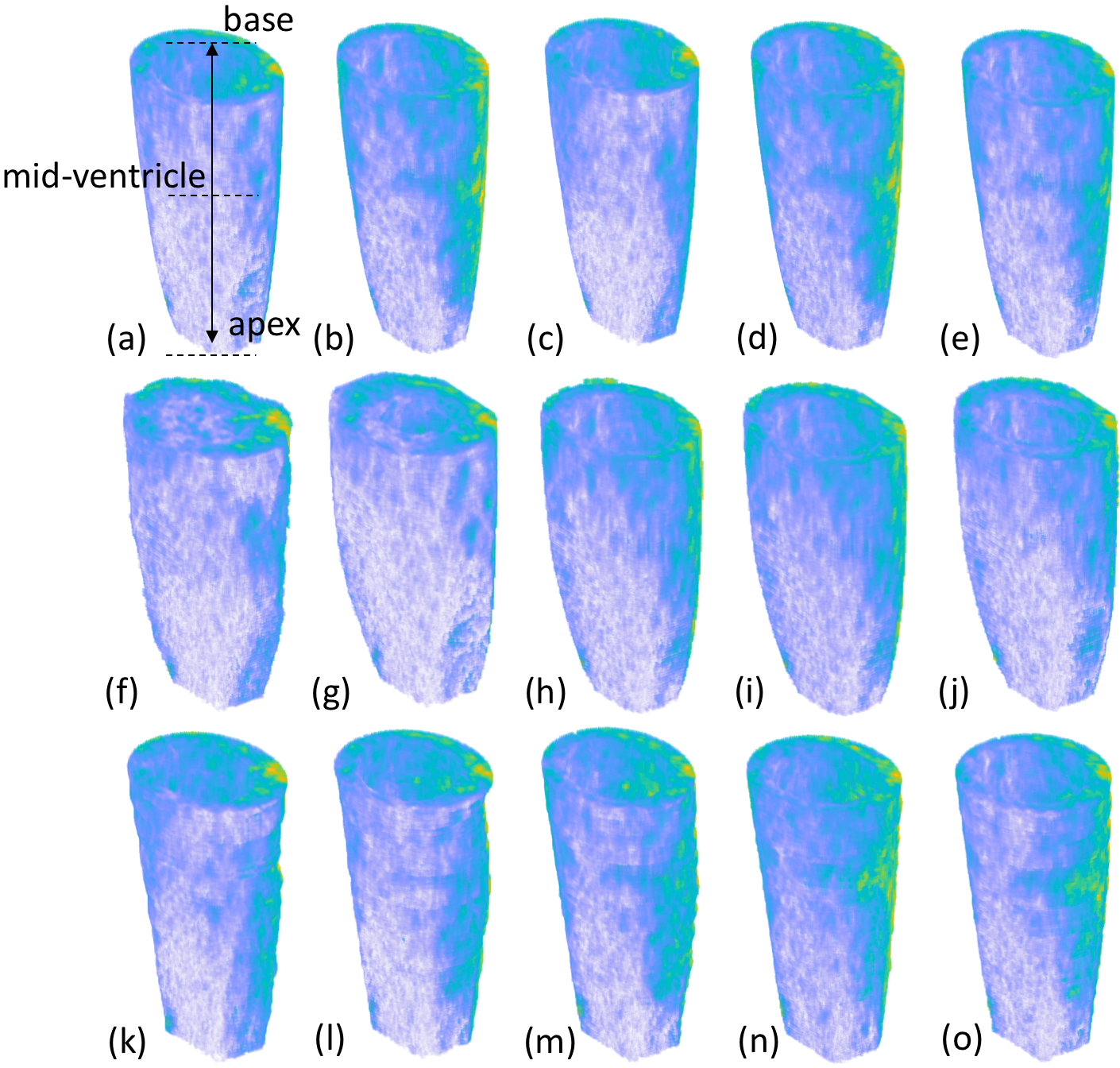}
 \end{center}
\caption{3D segmentation volumes of the left ventricle for 3D echocardiographic sequences illustrating [upper-row] ground truth for normal, ischemic-RCA, ischemic-LCX, ischemic-LADdist, and ischemic-LADprox; [mid-row] C-DAM method in \cite{hng14}; [bottom-row] proposed fully-automated fBm approach. (Video examples can be referred to in supplementary materials)}
\label{fig:LV_seg_3D}
\end{figure}

\begin{figure}[t]
 \begin{center}
 \includegraphics[scale=0.24]{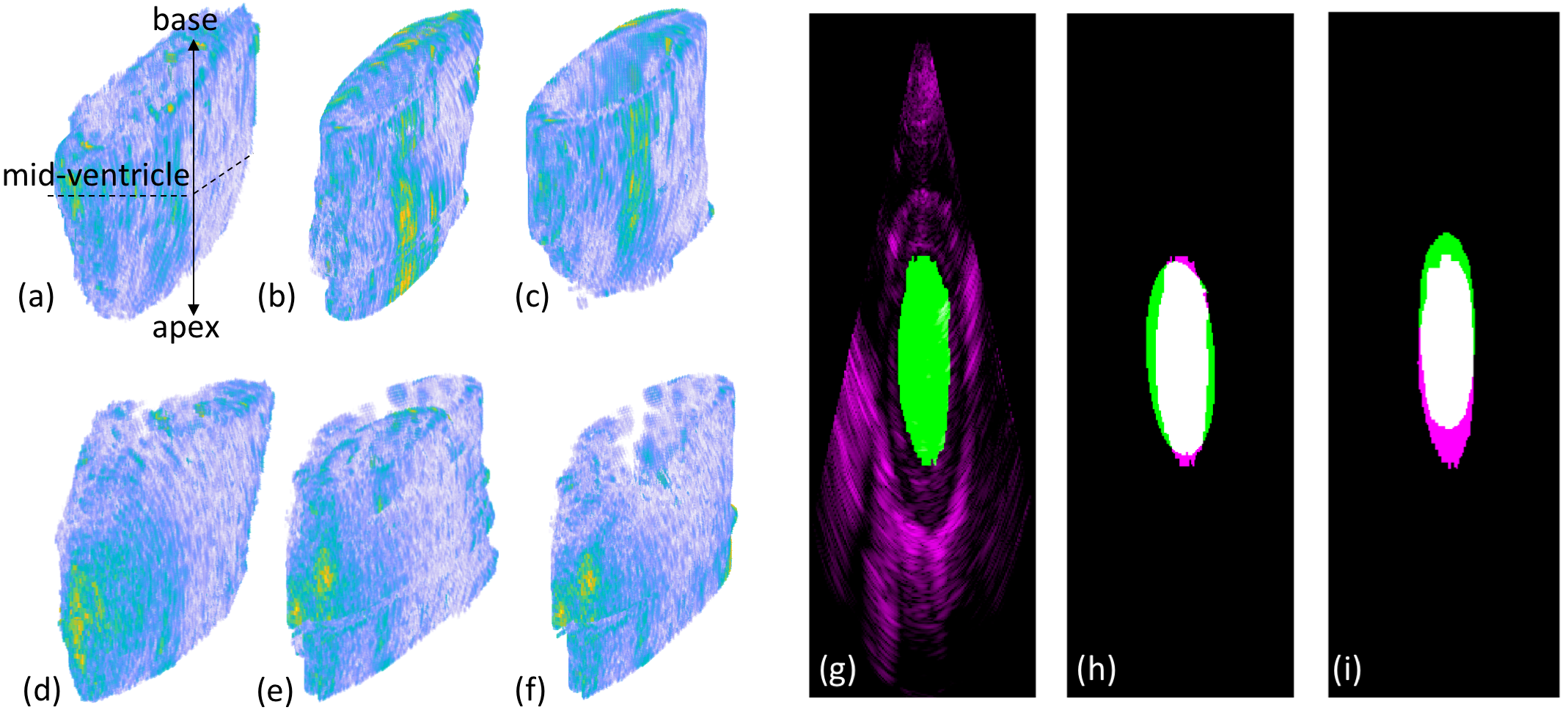}
 \end{center}
\caption{[Left] 3D segmentation of left ventricle for 2 canine subjects illustrating (a)-(c) manual, C-DAM, fBm in resting state condition, and (d)-(f) manual, C-DAM, fBm in stress state condition; [Right] comparison of endocardium segmentation accuracy of LV mid-ventricular cavity in resting state for (g)-(i) manual, fBm-based, C-DAM, respectively. (For qualitative interpretation of segmentation quality, figures (g)-(i) best viewed in color)}
\label{fig:LV_vol_cn_mid_slice}
\end{figure}

From a clinical perspective, one of the key properties of the fBm process is that it can exhibit persistence ($H > \tfrac{1}{2}$) or anti-persistence ($H < \tfrac{1}{2}$). Persistence is the property that the LV wall motion tends to be smooth, e.g. near to normal. Anti-persistence is the property that the relative stochastic process is very noisy, and hence LV wall motion trajectories tend to be heterogeneous. The latter case is an example of cardiac ischemia, such that displacements over one temporal or spatial interval are partially cancelled out by displacements over another time interval. From Table \ref{table:endo_class}, the fBm segmentation method showed improved performance for nearly all tested LV endocardium and epicardium deformation conditions except for the ischemic--LADdist condition. The occlusion of the distal and side branches could have a negative side-effect on LV myocardial function, introducing heterogeneity in the LV wall motion velocity pattern \cite{azv96}. Therefore the spatio-temporal structure of speckle patterns could be obscured or difficult to discern; affecting the fBm stochastic modelling for LADdist segmentation. Besides the possible presence of extensive epicardial fat may also complicate the LV segmentation. The endocardium of ischemic--LADprox showed nearly equivalent performance in both methods. For practical relevance, results are also demonstrated visually in Fig.~\ref{fig:LV_seg_3D} and Fig.~\ref{fig:LV_vol_cn_mid_slice}. Improved delineation of the epicardium and endocardium boundaries can be seen, especially for the LV base slice.

In the ellipsoidal modelling phase, where the fBm segmented image is combined with the corresponding shape information, the relative difference in magnitude of the eigenvalues $w$ and $l$ are an indication of the eccentricity of the LV. This gives an elliptical ventricle with a cone-shaped apex rather than having a round cylindrical-shape. It is known that the contraction of the LV cavity is less symmetrical because of the systolic increase in wall thickness \cite{kds95}. A way to view the LV is to consider it as a composite of adjoining structures; therefore a circular or elliptical shape at one point may not represent its entire structure. This can be attributed to the heterogeneity between the major axis diameter and cross-sectional area for the different regions of the LV during contraction, which is associated with the descent of the base and rotation of the apex. As the ventricular shape changes from an elliptical to a more spherical form, the assumption of a uniform structure of LV that is localized spatially would be misleading. The 3D spatio-temporal imaging would be advantageous for assessing the LV shape and size change at specific regions of interests. The purpose of the fBm-based method is to account for changes in shape from circular to elliptical as shown in Fig.~\ref{fig:LV_seg_3D}, where the difference in the major axis diameter from the apex to the base of the LV is refined by image moments. Nevertheless, other geometric primitives could be used, as well as deformable surfaces for improving LV boundary refinement of the fBm-based segmentation method.

Methods relying on evolving surfaces, e.g. \cite{hng14}, can allow for flexible topology changes and does not assume \textit{a priori} knowledge of object's shape; however, they cannot effectively segment surfaces that break apart or intersect -- which is a common condition in ultrasound imaging. Ultrasound image segmentation is challenging due to the inherent speckle and presence of artifacts such as shadows, attenuation and signal dropout. This often leads to missing edges, making it difficult for such methods to deal with structure discontinuity. Results in this work support that irregular structure characterization by locally investigating self-similarity patterns appearing at different scales using the fBm process can better quantify the heterogeneity in LV wall motion. The improved localization in both the temporal and spatial domains gives the fBm approach the advantage of accurate segmentation for fine details, and possible compensation of edge disconnection. Moreover, the non-stationarity of the fBm process integrates well with the stochastic nature of ultrasound echoes. 

Finally, due to the nature of how the ultrasound volume is segmented -- i.e. slice-based approach, the current implementation is not free of several limitations: 1) A disc-like structure is usually assumed in magnetic resonance (MR) data due to the imaging limitations for cine MR images. However, this approach might hamper the segmentation of a closed myocardium -- since true 3D ultrasound segmentation algorithms as in \cite{brd16} consider the LV as a closed structure, although not always assumed as such for the basal side, but typically for the apex; 2) Previous work has reported that different 3D echocardiography software packages show variability when used in predicting response to cardiac resynchronization therapy \cite{aly12}, therefore comparing performance with similar commercial software packages would be interesting to investigate; 3) Performing the fBm segmentation based on the radio-frequency envelope detected echos may better reflect the intrinsic properties of the myocardium tissue \cite{alk16, alk16b}, and assist in overcoming issues related to ultrasound B-mode settings.

\section{Conclusion}
\label{conclude}
This work presents a novel method for improving LV segmentation by addressing the problem of speckle pattern heterogeneity, where a) the fBm classification-based segmentation method relies naturally on the spatio-temporal dependencies of the local features; b) the 3D sequence of Hurst indices, which was used to derive fractal dimension volume maps, are invariant to intensity transformations; c) global information about the LV shape using second-order moments complements the local characterization of the fBm process. Both local and global boundary information about the LV shape boundaries was captured with improved precision. 

\appendix {}
\label{appendix}
\subsection{Covariance of fractional Brownian motion}
Given a Gaussian process that is characterized by associated $Hurst$ parameter $H \in (0, 1)$,
then it follows that the covariance function is given by
\begin{gather}
    \label{eqn:cov_fun}
    \rho(s,t) = \mathbb{E}(B_{H}(s) B_{H}(t)) \nonumber\\= \frac{1}{2}\big[|t|^{2H} + |s|^{2H} - |t - s|^{2H}\big], \quad \forall \enskip s,t \in \mathbb{R}^N,
\end{gather}

\noindent for $0 < s \leq t$, where $\mathbb{E}$ denotes the expectation operator with respect to probability space, and $|t|$ is the Euclidean norm of $t \in \mathbb{R}^N$. 

\subsection{Discrete fractional Brownian motion}
A discrete-time representation of fBm can be obtained by sampling the continuous-time fBm. When approximating (\ref{eqn:cont_fBm}) by sums, the first integral should be truncated, say at -$b$. The approximation $B_{H}(n)$ is for $n = 1,\dots,N$ given by

\begin{gather}
    \label{eqn:discrete_fBm}
    \tilde{B}_{H}(n) = C_{H} \bigg(\sum_{k=-b}^{0}[(n-k)^{H - 1/2} - (-k)^{H - 1/2}]B_{1}(k) \nonumber\\+ \sum_{k=0}^{n}(n-k)^{H - 1/2}B_{2}(k)\bigg),
\end{gather}

\noindent where $B_1$ with respect to $B_2$ are mutually independent vectors.% of $b + 1$ with respect to $N + 1$ i.i.d.\enskip standard normal variables.

\subsection{Fractal dimension estimation}
\label{FD_est}
There are several ways to estimate the fractal dimension of a stochastic process modeled by a fBm \cite{alk17}. All of them are based on the formula
\begin{equation}
    \label{}
    \mathbb{E}\big[(\tilde{B}_{H}(n+l) - \tilde{B}_{H}(n))^2 \big]
\end{equation}
for the variogram of fBm, which follows from (\ref{eqn:cov_fun}). The fBm and power-law variogram fits were used to estimate $H$ as a measure of self-similarity. For the discrete-time fBm process $\tilde{B}_{H}(n)$ defined in (\ref{eqn:discrete_fBm}), the following properties hold \cite{dch97}:

Property 1: The mean of the fBm increments, which are $l$ samples apart, is zero, i.e.,
\begin{equation}
\label{}
\mathbb{E}\big[\tilde{B}_{H}(n + l) - \tilde{B}_{H}(n)\big] = 0.
\end{equation}

The covariance of the fBm increments, $l$ samples apart, is given by
\begin{gather}
\gamma(l,k) = \mathbb{E}\big[(\tilde{B}_{H}(n + l) - \tilde{B}_{H}(n)) (\tilde{B}_{H}(n + l + k) \nonumber\\- \tilde{B}_{H}(n + k))\big] \nonumber\\
= \frac{\sigma^2}{2}\big[|k - l|^{2H} + |k + l|^{2H} - 2|k|^{2H}\big].\label{eqn:covmat}
\end{gather}

When $k= 0$ and $l = 1$, the variance of the unit increments is $\sigma^2$, which is the variance of independent identically distributed samples. From (\ref{eqn:covmat}), the following two properties of fBm can be obtained.

Property 2: For unit increments $l = 1$, the covariance of the increments becomes
\begin{equation*}
\gamma(k) = \frac{\sigma^2}{2}\big[|k - 1|^{2H} + |k + 1|^{2H} - 2|k|^{2H}\big].
\end{equation*}

Property 3: The variance of the increments when $k = 0$ is,
\begin{equation*}
\sigma^{2}_{H}(l) = \sigma^{2}|l|^{2H}.
\end{equation*}

The self-similarity characteristic of fBm can be recognized in property 3 with a scale transformation
\begin{gather*}
\sigma^{2}_{H}(rl) = \sigma^{2}|rl|^{2H}\\
= r^{2H}\sigma^2|l|^{2H}\\
= r^{2H}\sigma^2_{H} l,
\end{gather*}
\noindent where $r > 0$. Such a result indicates that fBm is statistically indistinguishable under a scale transformation and implies scale-invariance.
Combining property 1 and 3 to obtain
\begin{equation}
    \label{}
    \mathbb{E}\bigg[(\tilde{B}_{H}(n+l)-\tilde{B}_{H}(n))^2 \bigg] = \sigma^2 l^{2H},
\end{equation}

which is generally expressed in the form
\begin{equation}
    \label{eqn:var_fBm_com}
    \mathbb{E}\bigg[|\tilde{B}_{H}(n+l)-\tilde{B}_{H}(n)| \bigg] = c|l|^{H},
\end{equation}
where $c$ is proportional to the standard deviation $\sigma$.
Taking the logarithm of (\ref{eqn:var_fBm_com}) yields the linear equation
\begin{equation}
    \label{eqn:var_fBm_lq}
   \log \mathbb{E}\bigg[|\tilde{B}_{H}(n+l)-\tilde{B}_{H}(n)| \bigg] = H\log|l| + \log c.
\end{equation}

\ifCLASSOPTIONcompsoc
  % The Computer Society usually uses the plural form
  \section*{Acknowledgments}
\else
  % regular IEEE prefers the singular form
  \section*{Acknowledgment}
\fi

The author would like to thank Dr. Albert Sinusas's lab for provision of the canine dataset, Dr. James Duncan for very helpful discussion and making available the C-DAM method for comparison, and Allen Lu and Dr. Nripesh Parajuli for assisting with the delineation of the ground truth. This work was supported by Fulbright Scholarship Grant No. G-1-00005.

% Can use something like this to put references on a page
% by themselves when using endfloat and the captionsoff option.
\ifCLASSOPTIONcaptionsoff
  \newpage
\fi

\bibliographystyle{bibliographies/IEEEtran}
\bibliography{bibliographies/mybibfile.bib}

% insert where needed to balance the two columns on the last page with
% biographies
%\newpage

% You can push biographies down or up by placing
% a \vfill before or after them. The appropriate
% use of \vfill depends on what kind of text is
% on the last page and whether or not the columns
% are being equalized.

%\vfill

% Can be used to pull up biographies so that the bottom of the last one
% is flush with the other column.
%\enlargethispage{-5in}

% that's all folks
\end{document}